\documentclass{PoS}
\usepackage{graphicx}
\PoS{PoS(WC2004)034}

\title{Dilatonic, Chiral Cosmic Strings}

\ShortTitle{Dilatonic, Chiral Cosmic Strings}

\author{Vanessa C. Andrade \\
        Departamento de F\'{\i}sica, Univ. de Bras\'{\i}lia, Brazil\\
        E-mail: \email{andrade@fis.unb.br}}

\author{\speaker{André Luiz Naves de Oliveira}\\
Departamento de F\'{\i}sica, Univ. de Bras\'{\i}lia, Brazil\\
        E-mail: \email{ andreo@fis.unb.br}}

\author{Maria Em\'{\i}lia X. Guimar\~aes\\
        Departamento de Matem\'atica, Univ. de Bras\'{\i}lia,
        Brazil\\
        E-mail: \email{marg@unb.br}}

\abstract{In this work, we deal with the chiral string model for
which the world-sheet current is null in the framework of a
scalar-tensor gravity. Our main goal is to analyse the impact of
such a current on the gravitational macroscopic effects. For the
purpose of this analysis, we first study the gravitational
properties of the spacetime generated by this string in the
presence of a dilaton field. Then, we carry out an investigation
of the mechanism of formation and evolution of wakes in this
framework, showing the explicit contribution of the chirality to
this effect.}

\FullConference{Fourth International Winter Conference on Mathematical Methods in Physics\\
                 09 - 13 August 2004\\
                 Centro Brasileiro de Pesquisas Fisicas (CBPF/MCT), Rio de Janeiro, Brazil}

\begin{document}

\section{Introduction}

The present work deals with the case of a chiral string model in
the framework of a scalar-tensor gravity. In particular, we are
interested in the macroscopic description of the effect in cosmic
strings of mechanisms of the various (fermionic and bosonic
superconducting) kinds originally proposed by Witten~\cite{Wi85}.
Whilst considering current-carrying cosmic vortex, we will also
add scalar-tensor corrections on its metric. The motivation relies
on the fact that  theoretically, the possibility that gravity
might not be fundamentally described by purely tensorial theory in
four dimensions is growing in importance. This is in part a
consequence of superstring theory, which is consistent in ten
dimensions (or M-theory in eleven dimensions), but also the more
phenomenological recent developments of ``braneworld'' scenarios
\cite{ADD,RS} have motivated the study of other gravitational
theories in four-dimensions. In fact, the origin of the
(gravitational) scalar field can be many: the scalar field arising
from the size of the compactified internal space in the
Kaluza-Klein theory; the zero mode (dilaton field) described by a
symmetric second-rank tensor behaving as space-time metric at low
energy level in the closed string theory; the scalar field in a
brane world scenario; and more\footnote{For a deeper discussion on
this subject, we refer the reader to the book of Y. Fujii and K.
Maeda~\cite{maeda}.}. In any case, clearly, it is expected that
both the small scale implications of alternative theories of
gravity \cite{KKC} and the large scale implications of modified
gravity \cite{MNG} of such scenarios have a direct impact for
astrophysics and cosmology.

The combination of both conducting properties of a string and the
modified gravity (in particular, the scalar-tensor models) is, of
course, of real interest in astrophysics and cosmology and has
been treated elsewhere \cite{cristine1, andre1, andre2}. The most
noticeable effect was point out by Peter, Guimar\~aes and Andrade
\cite{vanessa} in which the (gravitational) scalar
field\footnote{From now on, we will call this field generically as
``dilaton".} behaves as a winding phase along the string, thereby
generating a neutral current kind of effect by raising the
degeneracy between the eigenvalues of the stress-energy tensor.

Our purpose in this paper is twofold. We first present the metric
generated by a chiral cosmic string in the scalar-tensor gravity.
This metric is obtained straightforwardly by an algorithm
developed in the reference \cite{felipe}. The second and main goal
of this work is to explore how this null current - chirality -
affects the motion of particles in this spacetime. Namely, we
analyze the formation and evolution of wakes in this model. We
anticipate that the main conclusion of this paper is to show that,
up to first order in a parameter of the underlying theory, the
mechanism of accretion is not sensitive to the ``conducting
properties" of this model, the result being qualitatively  very
similar to the mechanism by a neutral string.

This work is outlined as follows. In section 2, we present the
gravitational field of a scalar-tensor chiral string and we
consider the mechanism of formation and evolution of wakes in this
framework by means of the Zel'dovich approximation. In section 3,
we summarize the main conclusions and compare our results with the
case of a neutral cosmic string in general relativity
\cite{vachaspati} and with the cases of a neutral \cite{sandra}
and a chiral \cite{abdalla} in scalar-tensor gravity.

\section{The scalar-tensor chiral string and the wake formation}

\subsection{The gravitational field}

In this work, we will modify gravity within a class of
scalar-tensor theories such that the action is written as:
\begin{equation}
\label{acao1} {\cal S}= \frac{1}{16\pi} \int d^4 x
\sqrt{-\tilde{g}} \left[ \tilde{R}\tilde{\Phi} -
\frac{\omega(\tilde{\Phi})}{\tilde{\Phi}}
\partial^{\mu}\tilde{\Phi}\partial_{\mu}\tilde{\Phi}\right]
+ {\cal S}_{m}[\Psi_m , \tilde{g}_{\mu\nu}] \,\, ,
\end{equation}
$\tilde{g}_{\mu\nu}$ is the physical metric which contains both
scalar and tensor degrees of freedom, $\tilde{R}$ is the curvature
scalar associated to it and ${\cal S}_{m}$ is the action for
general matter fields which, by now, is left arbitrary.

By varying action (\ref{acao1}) with respect to the metric
$\tilde{g}_{\mu\nu}$ and to the scalar field $\tilde{\Phi}$ we
obtain the modified Einstein equations and a wave equation for
$\tilde{\Phi}$:
\begin{eqnarray}
\label{equacao1} \tilde{R}_{\mu\nu} -
\frac{1}{2}\tilde{g}_{\mu\nu}\tilde{R} & = &
\frac{8\pi}{\tilde{\Phi}}\tilde{T}_{\mu\nu} +
\frac{\omega(\tilde{\Phi})}{\tilde{\Phi}}\left[\partial_{\mu}\tilde{\Phi}
\partial_{\nu}\tilde{\Phi} -\frac{1}{2}\tilde{g}_{\mu\nu}\partial^{\alpha}
\tilde{\Phi}\partial_{\alpha}\tilde{\Phi} \right] \nonumber \\
& & +  \frac{1}{\tilde{\Phi}}\left(\nabla_{\nu}\tilde{\Phi}_{,\mu}
-
\tilde{g}_{\mu\nu}\Box_{\tilde{g}}\tilde{\Phi} \right) , \nonumber \\
\Box_{\tilde{g}}\tilde{\Phi} & = & \frac{1}{2\omega(\tilde{\Phi})
+ 3} \left[ 8\pi \tilde{T} - \frac{d\omega}{d\tilde{\Phi}}
\partial_{\mu}\tilde{\Phi}\partial^{\mu}\tilde{\Phi} \right], \\
\nabla_{\mu}\tilde{T}^{\mu}_{\nu} & = & 0 \,\, ,
\end{eqnarray}
where
\[
\tilde{T}_{\mu\nu} = \frac{2}{\sqrt{-\tilde{g}}} \frac{\delta
{\cal S}_m}{\delta \tilde{g}^{\mu\nu}} \,\, ,
\]
is the energy-momentum tensor of the matter content and $\tilde{T}
\equiv \tilde{T}^{\mu}_{\mu}$ is its trace. Clearly, there is an
equivalence between the scalar-tensor action (\ref{acao1}) and the
general relativity case, though this feature is not generally
true.  If $\tilde{T}$ vanishes and $\tilde{\Phi}$ is a constant,
equations (\ref{equacao1}) reduce to the usual Einstein's
equations if we identify a gravitational constant $G^*$ with the
inverse of the scalar field, e.g., $G^* = 1/\tilde{\Phi}$. Hence,
any exact solution of Einstein's equations with a trace-free
matter source will also be a particular exact solution of the
scalar-field with $\tilde{\Phi}$ constant. Of course, this
particular solution will not be the general solution for the
matter content. {\it En passant}, we mention that, when the scalar
field kinetic term is absent in (\ref{acao1}), these theories are
equivalent to the $1/R$ gravity theories. In this case, the scalar
field is {\it not} an independent dynamical degree of freedom and,
among many consequences, these theories are in conflict with solar
system observations \cite{chiba} and are not suitable to describe
the cosmic acceleration \cite{flanagan}.

If we make a  conformal deformation
\begin{equation}
\label{equacao2} \tilde{g}_{\mu\nu} = A^2(\phi) g_{\mu\nu} \,\, ,
\end{equation}
and  a redefinition of the quantities
\begin{equation}
G^* A^2(\phi) = \frac{1}{\tilde{\Phi}} \,\, ,
\end{equation}
$G^*$ is a ``bare" gravitational constant, and
\begin{equation}
\alpha(\phi) = \frac{ \partial{\ln A(\phi)}}{\partial \phi} =
\frac{1}{(2\omega(\tilde{\Phi}) + 3)^{1/2}} \,\, ,
\end{equation}
which can be interpreted as the (field-dependent) coupling
strength between matter and scalar field, then (\ref{acao1}) can
be re-written in the Einstein (also called conformal) frame in
which the kinematic terms of tensor and scalar do not mix:
\begin{equation}
\label{acao2} {\cal S} = \frac{1}{16\pi G^*} \int \sqrt{g} \left[
R - 2g^{\mu\nu} \partial_{\mu}\phi \partial_{\nu}\phi \right] +
{\cal S}_m [\Psi_m , A^2(\phi)g_{\mu\nu}] \,\, ,
\end{equation}
where $g_{\mu\nu}$ is a pure rank-2 metric tensor and $R$ is the
curvature scalar associated to it.

In the conformal frame, eqs. (\ref{equacao1}) are written in a
more convenient form:
\begin{eqnarray}
\label{equacao3} R_{\mu\nu} - \frac{1}{2}g_{\mu\nu}R & = & 8\pi
G^* T_{\mu\nu} + 2\partial_{\mu}\phi\partial_{\nu}\phi -
g_{\mu\nu}g^{\alpha\beta}\partial_{\alpha}\phi\partial_{\beta}\phi \,\, , \nonumber \\
\Box_{g}\phi & = & - 4\pi G^* \alpha(\phi) T \,\, .
\end{eqnarray}

It has been shown in \cite{felipe} that we can relate the metric
in the scalar-tensor gravities with the metric in Einstein's
gravity for the same matter distribution in the weak-field
approximation regime. For a matter of consistency, we will
summarize this method below.

Let us expand the fields to first order in the parameter $G_0 =
G^*A^2(\phi_0)$, we then obtain
\begin{eqnarray}
\label{ee}
g_{\mu\nu} & = & \eta_{\mu\nu} + h_{\mu\nu} \,\, , \nonumber \\
\phi & = & \phi_0 + \phi_{(1)} \,\, , \\
A(\phi) & = & A(\phi_0)[ 1 + \alpha(\phi_0)\phi_{(1)}] \,\, , \nonumber \\
T^{\mu}_{\nu} & = & T^{\mu}_{(0)\nu} + T^{\mu}_{(1)\nu} \,\, .
\nonumber
\end{eqnarray}
Therefore, eqs. (\ref{equacao3}) reduce to:
\begin{eqnarray}
\label{li} \nabla^2 h_{\mu\nu} & = & 16\pi G^* (T^{(0)}_{\mu\nu} -
\frac{1}{2} \eta_{\mu\nu}T_{(0)}) \,\, , \\
\nabla^2 \phi_{(1)} & = & 4\pi G^* \alpha(\phi_0)T_{(0)} \,\, .
\nonumber
\end{eqnarray}

In this approximation, $T^{(0)}_{\mu\nu}$ is the energy-momentum
tensor at zeroth-order in the conformal frame. Its relation to the
(physical) energy-momentum tensor at zeroth-order in the
Jordan-Fierz frame is given by
$T^{(0)}_{\mu\nu}=A^2(\phi_0)\tilde{T}^{(0)}_{\mu\nu}$. In this
way, the first equation in the system (\ref{li}) is the Einstein's
equation in the weak-field approximation regime.

Once we have $\tilde{g}_{\mu\nu} = A^2(\phi)[\eta_{\mu\nu} +
h_{\mu\nu}]$, we can then use the approximation (\ref{ee}). We
obtain:
\begin{equation}
\tilde{g}_{\mu\nu} = A^2(\phi_0)[ 1 +
2\alpha(\phi_0)\phi_{(1)}][\eta_{\mu\nu} + h_{\mu\nu}] \,\, .
\end{equation}

We can apply now this algorithm to the case of the chiral string.
In the weak field approximation, the method described above is,
schematically
\begin{equation}
ds^2_{ST}=[1+2\alpha(\phi_0) \phi_{(1)}]ds^2_{GR} \,\, ,
\end{equation}
where $ds^2_{ST}$ is the line element in the scalar-tensor theory,
and $ds^2_{GR}$ is the line element in general relativity and
$\phi_{(1)}$ is the solution of the equation
\begin{equation}
\nabla^2\phi_{(1)}= 4\pi G^*\alpha(\phi_0) T_{(0)} \,\, .
\end{equation}
In the ref. \cite{steer1}, the metric and the energy-momentum
tensor for the chiral cosmic string in general relativity for the
weak field regime were obtained:
\begin{eqnarray*}
ds^2&=&dt^2\left[1+X(r,k)\right]-dz^2\left[1-X(r,k)\right]-dr^2 \nonumber\\
&+&\left[1-4G^* m^2\right]^2r^2d\theta^2-2X(r,k)dt dz \,\, ,
\end{eqnarray*}
where $0 \leq k \leq 1$ is the chiral factor which characterizes
the state of the string: $k=0$ corresponds to the maximal charged
strings and $k=1$ corresponds to the ordinary, neutral string. It
can happen that $k$ is not constant, but we will not deal with
this case here,
\begin{eqnarray*}
T_{(0)}^{\mu\nu}=\frac{2m^2}{1+k}\left(
    \begin{array}{cccc}
    1 & 0& 0& \frac{1-k}{2} \\
    0&0&0&0 \\
    0&0&0&0 \\
    \frac{1-k}{2}& 0&0& -k
    \end{array} \right) \delta(x)\delta(y) \,\, ,
    \end{eqnarray*}
where
$X(r,k)=8G^*m^2\frac{1-k}{1+k}\ln\left(\frac{r}{r_0}\right)$. With
this, we find the trace of the energy-momentum tensor ($T_{(0)}=2
A^2(\phi_0)m^2\delta(x)\delta(y)$). Using the result
$\nabla^2\ln\left(\frac{r}{r_0}\right)=2\pi\delta(x)\delta(y)$ we
have
\begin{equation}
\nabla^2\phi_{(1)}= 8\pi m^2G_0\alpha(\phi_0)\delta(x)\delta(y)
\rightarrow
\phi_{(1)}=4m^2G_0\alpha(\phi_0)\ln\left(\frac{r}{r_0}\right) \,\,
.
\end{equation}
Therefore, the metric for the scalar-tensor chiral string is, to
first order in $G_0$
\begin{eqnarray}
\label{chiralmetric}
ds^2_{ST}&=&\left(1+8\alpha(\phi_0)^2m^2G_0\ln\left(\frac{r}{r_0}\right)\right)
[(1+8G_0 m^2\frac{1-k}{1+k}\ln\left(\frac{r}{r_0}\right))dt^2\nonumber \\
&-&dr^2-(1-4G_0m^2)r^2d\theta^2 - (1-8G_0
m^2\frac{1-k}{1+k}\ln\left(\frac{r}{r_0}\right))dz^2 \nonumber
\\
& - & 2 (8G_0 m^2\frac{1-k}{1+k}\ln\left(\frac{r}{r_0}\right))dz
dt] \,\, .
\end{eqnarray}
Consequently, associated with the metric (\ref{chiralmetric}), we
have a gravitational force acting on a test particle of mass $M$:
\[
F =
-\frac{4M}{r}\left(\alpha(\phi_0)^2m^2+\left(\frac{1-k}{1+k}\right)m^2\right)G_0
\,\, .
\]
\subsection{The Zel'dovich approximation and the accretion problem}

To determine the displacement of a particle flowing past a cosmic
string to study in the next section the formation and evolution of
wakes, we will use the Zel'dovich approximation
\cite{zeldovich1970}.  To start with, we first compute the
velocity perturbation of massive particles moving past the string.
If we consider that the string is moving with normal velocity
$v_s$ through matter, the velocity perturbation can be
calculated\footnote{All expansions in the parameter $G_0$ were
performed using the algebraic computational programme MAPLE.} here
with the help of the gravitational force due to metric
(\ref{chiralmetric}):
\begin{equation}
\label{vi} u_i(r)=8\pi G_0 m^2v_s\gamma+ \frac{4\pi
G_0\alpha(\phi_0)^2m^2}{v_s\gamma} +
\left(\frac{1-k}{1+k}\right)\frac{4\pi G_0m^2}{v_s\gamma} \,\, ,
\end{equation}
with $\gamma = (1-v_s^2)^{-1/2}$. The first term in (\ref{vi}) is
equivalent to the relative velocity of particles flowing past a
string in general relativity. The other terms come as a
consequence of the scalar-tensor coupling of the gravitational
interaction and the chirality properties of the string.

Let us suppose now that the wake was formed at $t_i > t_{eq}$. The
physical trajectory of a dark particle can be written as
\begin{equation}
\label{traj} h(\vec{x}, t) = a(t) [ \vec{x} + \psi(\vec{x}, t)]
\,\, ,
\end{equation}
where $\vec{x}$ is the unperturbed comoving position of the
particle and  $\psi(\vec{x}, t)$ is the comoving displacement
developed as a consequence of the gravitational attraction induced
by the wake on the particle. Suppose, for simplification, that the
wake is perpendicular to the $x$-axis (assuming that $dz=0$ in the
metric (\ref{chiralmetric}) and $ r = \sqrt{x^2 + y^2}$) in such a
way that the only non-vanishing component of $\psi$ is $\psi_x$.
Therefore, the equation of motion for a dark particle in the
Newtonian limit is
\begin{equation}
\label{newton} \ddot{h} =  - \nabla_h \Phi \,\, ,
\end{equation}
where the Newtonian potential $\Phi$ satisfies the Poisson
equation
\begin{equation}
\label{poisson} \nabla_h^2 \Phi = 4\pi G_0 \rho \,\, ,
\end{equation}
where $\rho(t)$ is the dark matter density in a cold dark matter
universe. For a flat universe in the matter-dominated era, $a(t)
\sim t^{2/3}$. Therefore, the linearised equation for $\psi_x$ is
\begin{equation}
\label{psi} \ddot{\psi} + \frac{4}{3t}\dot{\psi} -
\frac{2}{3t^2}\psi = 0 \,\, ,
\end{equation}
with appropriated initial conditions: $\psi(t_i) = 0$ and
$\dot\psi(t_i) = -u_i$. Eq. (\ref{psi}) is the Euler equation
whose solution is easily found
\[
\psi(x,t) = \frac{3}{5}\left[\frac{u_i t_i^2}{t} - u_i t_i
\left(\frac{t}{t_i}\right)^{2/3}\right] \,\, .
\]

Calculating the comoving coordinate $x(t)$ using the fact that
$\dot{h} = 0$ in the ``turn around"\footnote{The moment when the
dark particle stops expanding with the Hubble flow and starts to
collapse onto the wake.}, we get
\begin{equation}
\label{comoving} x(t) = - \frac{6}{5} \left[ \frac{u_i t_i^2}{t} -
u_i t_i \left(\frac{t}{t_i}\right)^{2/3}\right] \,\, .
\end{equation}
The thickness and the wake's superficial density are,
respectively:
\begin{eqnarray}
\label{d} d(t)&\approx
&2x(t)\left(\frac{t}{t_i}\right)^{\frac{2}{3}}
=\frac{12}{5}u_it_i \left(\frac{t^{\frac{4}{3}}}
{t_i^{\frac{4}{3}}}\right)=\frac{12}{5}( 8\pi G_0 m^2v_s\gamma \\
&+& \frac{4\pi G_0\alpha(\phi_0)^2m^2}{v_s\gamma} +
\left(\frac{1-k}{1+k}\right)\frac{4\pi G_0m^2}{v_s\gamma}) \,\, ,
\nonumber
\end{eqnarray}
and
\begin{eqnarray}
\label{sigma} \sigma(t)&\approx &\rho (t) d(t)= \frac{2u_i}{5\pi
G_0 t}
(\frac{t}{t_i})^{\frac{1}{3}} \nonumber \\
&=&( \frac{16}{5}\frac{m^2v_s\gamma}{t}
+\frac{8}{5}\frac{\alpha(\phi_0)^2m^2}{v_s\gamma t} + \frac{8}{5}
\left(\frac{1-k}{1+k}\right) \frac{m^2}{v_s\gamma t})
\left(\frac{t}{t_i}\right)^{\frac{1}{3}} \,\, .
\end{eqnarray}
We summarize our results in the next section.

\section{Main results}

Inclusion of a current in the internal structure of a cosmic
vortice can drastically change the predictions of such models in a
number of small and large-scale effects. In particular, models
with a timelike-type or a spacelike-type currents can present
divergences leading to unbounded gravitational effects
\cite{andre1,andre2}.

In this work, we have concentrated on the chiral string model for
which the world-sheet current is null in the framework of a
scalar-tensor gravity. Our main goal was to analyse the impact of
such a current on the gravitational macroscopic effects. For the
purpose of this analysis, we first studied the gravitational
properties of the spacetime generated by this string in the
presence of a dilaton field. Then, we carried out an investigation
of the mechanism of formation and evolution of wakes in this
framework, showing the explicit contribution of the chirality to
this effect.

Surprisingly enough, the mechanism of accretion is not sensitive
to the ``conducting properties" of this model, the results being
qualitatively  very similar to the mechanism by a neutral string,
up to first order in the parameter $G_0$. The divergences appear
at higher orders in this parameter.

\subsection*{Comparison with other cosmic string models}

\begin{figure}
\centering
\includegraphics[angle=270,scale=0.3]{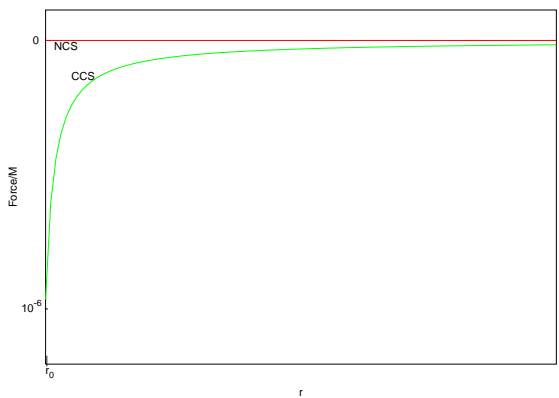}
\caption{Gravitational force for a neutral (NCS) and chiral
strings (CCS) in general relativity} \label{forcarg}
\end{figure}

\begin{figure}
\centering
\includegraphics[angle=270,scale=0.3]{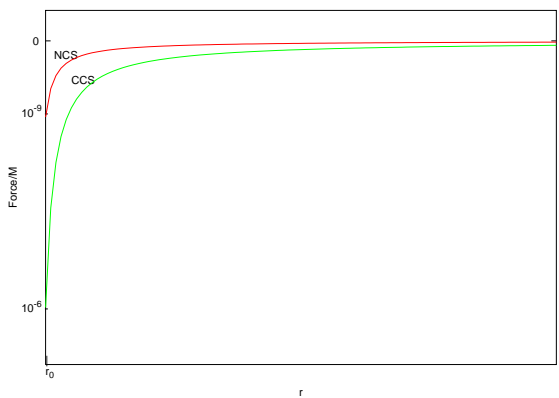}
\caption{Gravitational force for a neutral (NCS) and chiral (CCS)
strings in scalar-tensor theories} \label{forcast}
\end{figure}

\begin{itemize}

\item{\bf  Chiral strings in general relativity and scalar-tensor
theories}

It is interesting to observe that chiral string  models behave
qualitatively in the same way for both theories of gravity. In the
weak-field approximation, which is indeed the regime under
consideration, the scalar-tensor model has an additional factor of
order $10^{3}$ smaller than his general relativity partner. This
factor is introduced by the post-Newtonian parameter $\alpha_0$
which expresses the strength of the interaction between matter and
the dilaton field.

\item{\bf  Chiral and neutral strings in general relativity and
scalar-tensor theories}

To introduce chirality and dilatonic properties in a cosmic string
model definitely changes the behavior of the gravitational force
on test particles, as can be seen from the Figs. 1 and 2.
Nevertheless, the impact on the formation of wakes is neglegible
to first order in the parameter $G_0$.

\end{itemize}

\section*{Acknowledgements}

M. E. X. Guimar\~aes would like to thank the Conselho Nacional de
Desenvolvimento Cient\'{\i}fico e Tecnol\'ogico (CNPq) for a
support and A. L. Naves de Oliveira would like to thank the
Coordena\c{c}\~ao de Aperfei\c{c}oamento de Pessoal de N\'{\i}vel
Superior (CAPES/MEC) for a PhD grant. The authors would like to
thank FINATEC (UnB) for partial support.

\end{document}